
\input jnl.tex
\input reforder.tex

\vglue 0.5 truein

\title
{
Quantum Cosmology
}
\smallskip
\author
{Alexander Vilenkin}
\affil
{
Tufts Institute of Cosmology,
Department of Physics and Astronomy,
Tufts University, Medford, MA 02155
}


\body


If the cosmological evolution is followed back in time, we come to the
initial singularity where the classical equations of general relativity
break down.  This led many people to believe that in order to understand
what actually happened at the origin of the universe, we should treat the
universe quantum-mechanically and describe it by a wave function rather
than by a classical spacetime.  This quantum approach to cosmology was
originated by DeWitt\refto{dw} and Misner\refto{cm}, and after a somewhat
slow start has become
very popular in the last decade or so.  The picture that has emerged from
this line of development\refto{et, pif, avpl, gz, hh, lnc, avpr}
is that a small closed universe can spontaneously
nucleate out of ``nothing", where by ``nothing" I mean a state with no
classical space and time.  The cosmological wave function can be used to
calculate the probability distribution for the initial configurations of the
nucleating universes.  Once the universe nucleated, it is expected to go
through a period of inflation, which is a rapid (exponential) expansion
driven by the energy of a false vacuum.  The vacuum energy eventually
thermalized, inflation ends, and from then on the universe follows the
standard hot cosmological scenario.  Inflation is a necessary ingredient in
this kind of scheme, since it gives the only way to get from the tiny
nucleated universe to the large universe we live in today.

In this talk I would like to review where we stand in this programme.
Despite the large amount of work in quantum cosmology, we still do not have
a ``standard model", and I am sure that you would get a very different
picture if this talk were to be given by Stephen Hawking, Jim Hartle or
Jonathan Halliwell.  I think my view is close to that of Andrei Linde,
although we tend to emphasize different things.


\beginsection{A SIMPLE MODEL}

First I would like to illustrate how the nucleation of the universe can be
described in a very simple model.  I assume that the universe is
homogeneous and isotropic, so that it is described by the closed Robertson-
Walker metric
$$
ds^2 = dt^2 - a^2(t) d\Omega_3^2 \,. \eqno (1)
$$
(The universe should be closed, since otherwise its volume would be
infinite and the nucleation probability would be zero).  The scale factor
$a(t)$ satisfies the evolution equation
$$
{\dot a}^2 +1 = {8\pi G \over{3}} \rho a^2 \,. \eqno (2)
$$
The simplest inflationary model is the one in which $\rho$ is a constant vacuum
energy density, $\rho = \rho_v$.  Then the solution of (2) is the de Sitter
space,
$$
a(t) = H^{-1} \cosh (Ht) \,, \eqno (3)
$$
where
$$
H = (8\pi G \rho_v /3)^{1/2}  \,.
$$
The universe contracts at $t<0$, reaches the minimum radius $a = H^{-1}$ at $t
= 0$ and re-expands at $t > 0$.

This is similar to the behavior of a particle bouncing off a potential
barrier, with $a$ playing the role of particle coordinate.  Now, we know that
in quantum mechanics particles can not only bounce off, but can also tunnel
through potential barriers.  This suggests the possibility that the
negative-time part of the evolution in (3) may be absent, and that the
universe may instead tunnel from $a = 0$ directly to $a = H^{-1}$.  To see
whether this is
indeed the case, we should quantize our simple model.

The quantization amounts to replacing the momentum, $p_a = -a{\dot a}$,
conjugate to the
variable $a$ by an operator $-id/ da$.  Then, disregarding the factor-ordering
ambiguities, the evolution equation (2) yields the Schrodinger equation
$$
\left[ {d^2 \over{d a^2}} - U(a) \right] \psi (a) = 0 \,, \eqno
(5)
$$
where
$$
U(a) = a^2 (1 - H^2 a^2 ) \,. \eqno (6)
$$
The ``potential" $U(a)$ has the form of a barrier separating $a = 0$ and $a =
H^{-1}$, and it
is clear that Eq. (5) should have a tunneling solution.  This solution is
specified by requiring that $\psi$ has only an outgoing wave at $a \to \infty$.
 For $H << m_{\rm pl}$,
the wave function can be found using the semiclassical approximation and
can be used to calculate the ``tunneling probability"\refto{lnc, avpr}.
However, I shall
postpone the discussion of probabilities until we consider a more general
model.

\beginsection{WAVE FUNCTION OF THE UNIVERSE}

In the general case, the wave function of the universe is defined on
superspace, which is the space of all 3-geometrics $g_{ij}({\bf x})$ and
matter field
configurations $\varphi ({\bf x})$,
$$
\psi [g_{ij}({\bf x}), ~ \varphi ({\bf x})] \,. \eqno (7)
$$
$\psi$ satisfies the Wheeler-DeWitt (WDW) equation\refto{dw},
$$
{\cal H} \psi(g_{ij},\varphi) = 0 \,, \eqno (8)
$$
which can be thought of as expressing the fact that the energy of a closed
universe is equal to zero.  The WDW equation can be symbolically written in
the form
$$
(\nabla^2 - U)\psi = 0 \,, \eqno (9)
$$
which is similar to the Klein-Gordon equation.  Here, $\nabla$ is a functional
differential operator on superspace and the function $U(g_{ij}, \varphi)$ can
be called
``superpotential".

Quantum cosmology is based on quantum gravity and shares all of its
problems, in particular the uncontrollable infinities.  In addition it has
some extra problems which arise when one tries to quantize a closed
universe.  The first problem stems from the fact that $\psi$ is independent of
time.  This can be understood\refto{dw} in the sense that the wave function of
the
universe should describe everything, including the clocks which show time.
In other words, time should be defined instrinsically in terms of the
geometric or matter variables.  However, no general prescription has yet
been found that would give a function $t(g_{ij}, \varphi)$ that would be, in
some sense,
monotonic.  A related problem is the definition of probability.  Given a
wave function $\psi$, how can we calculate probabilities?  One can try to use
the conserved current\refto{dw, cm}
$$
J = i(\psi^* \nabla \psi - \psi \nabla \psi^*),~~~ \nabla \cdot J = 0 \,. \eqno
(10)
$$
The conservation is a useful property, since we want probability to be
conserved.  But one runs into the same problem as with Klein-Gordon
equation:  the probability defined in this way is not positive-definite.
Althouth we do not know how to solve these problems in general, they can
both be solved in the semiclassical domain.  In fact, it is possible that
this is all we need.

\beginsection{SEMICLASSICAL UNIVERSES}

Let us consider the situation when some of the variables $\{c\}$ describing the
universe behave classically, while the rest of the variables $\{q\}$ must be
treated quantum-mechanically.  Then the wave function of the universe can
be written as
$$
\psi = \Sigma A(c) e^{iS(c)} \chi (c,q) \equiv \Sigma \psi_c \chi \,, \eqno
(11)
$$
where the classical variables are described by the WKB wave functions $\psi_c =
Ae^{iS}$.
In the semiclassical approximation, $\nabla S$ is large, and substitution of
(11) into the WDW equation (9) yields
the Hamilton-Jacobi equation for $S(c)$,
$$
\nabla S \cdot \nabla S + U = 0 \,. \eqno (12)
$$
Each solution of (12) is a classical action describing a congruence of
classical trajectories (which are essentially the gradient curves of $S$).
Hence, a semiclassical wave function $\psi_c = Ae^{iS}$ describes an ensemble
of classical
universes evolving along the trajectories of $S(c)$.  A probability
distribution for these trajectories can be obtained using the conserved
current (10).  Since the variables $c$ behave classically, the probabilities
do not change in the course of evolution and can be thought of as
probabilities for various initial conditions.  The time variable $t$ can be
defined as any monotonic parameter along the trajectories, and it can be
shown\refto{dw, avint} that in this case the corresponding component of the
current $J$ is
non-negative, $J_t \geq 0$.  Moreover, one finds\refto{lr, halh, ban}, that the
``quantum" wave function $\chi$
satisfies the usual Schrodinger equation,
$$
i \partial \chi /\partial t = H_\chi \chi \,        \eqno (13)
$$
with an appropriate hamiltonian $H_\chi$.
Hence, all the familiar physics is recovered in the semiclassical regime.

This semiclassical interpretation of the wave function $\psi$ is valid to the
extent that the WKB approximation for $\psi_c$ is justified and the
interference
between different terms in (11) can be neglected.  Otherwise, time and
probability cannot be defined, suggesting that the wave function has no
meaningful interpretation.  In a universe where no object behaves
classically (that is, predictably), no clocks can be constructed, no
measurements can be made, and there is nothing to interpret.

Suppose for a moment that the cosmological wave function $\psi$ is known, the
semiclassical domain is identified, and the probability distribution for
the ensemble of universes is calculated.  How can we test this distribution
observationally?  Strictly speaking, one needs an observer who would survey
the universes in the ensemble.  We do not know how to get in touch with
such an observer and can make predictions only if we assume that we live in
a ``typical" universe.  If the probability distribution has a strong peak,
we make a prediction.

It should be noted that the semiclassical approach to time and probability
outlined in this section has been implemented only in the context of
minisuperspace, which includes only a finite number of degrees of freedom,
or perturbative superspace in which all but a few degrees of freedom are
treated as small perturbations.  An extension to the general case may be
non-trivial.  For an up to date discussion see the review articles
by Kuchar\refto{kk} and Isham\refto{ci}.

\beginsection{BOUNDARY CONDITIONS}

Thus, to explain the initial conditions of the universe, all we need to do
is find the wave function $\psi$ from the WDW equation (9).  However, as any
differential equation, it has an infinite number of solutions.  To get a
unique solution, one has to specify some boundary conditions for $\psi$.  In
ordinary quantum mechanics, the boundary conditions for the wave function
are determined by the physical setup external to the system under
consideration.  In quantum cosmology, there is nothing external to the
universe, and a boundary condition should be added to eq.(9) as an
independent physical law.

Several candidates for this law of boundary condition have been proposed.
One of these is the tunneling boundary condition\refto{avbc}, which was
inspired by the
picture of the universe tunneling from ``nothing".  It requires that at the
boundaries of superspace $\psi$ should include only outgoing waves.  It is not
clear whether or not incoming and outgoing waves can be rigorously defined
in general, but they certainly can be defined in the semiclassical
approximation.  As we discussed in the previous section, the WKB wave
function (11) describes a congruence of classical trajectories, and the
tunneling boundary condition requires that these trajectories can end, but
cannot begin at the boundaries.  The boundaries of superspace correspond
to singular geometries and matter
fields, and a typical cosmological trajectory will both begin and end at
the boundary.  The tunneling condition selects the trajectories which begin
at the ``barrier", where some components of  $\nabla S$ vanish and the
semiclassical approximation breaks down.

A different proposal has been made by Hartle and Hawking\refto{hh, haw},
who require that
the wave function $\psi (g_{ij}, \varphi)$ should be given by a Euclidean
(imaginary-time) path integral over
compact 4-geometries bounded by the 3-geometry $g_{ij}$ with the field
configuration $\varphi$,
$$
\psi = \int [dg_{ij}] [d\varphi] e^{-S_E} \,. \eqno (14)
$$
This proposal is motivated by mathematical elegance and simplicity.
However, as it stands, the integral (14) is badly divergent, because the
Euclidean action $S_E$ is not positive-definite.  Attempts to define it by
analytic continuation were unsuccessful\refto{harhal}, but again the
Hartle-Hawking wave
function can be well defined in the semiclassical approximation.
Linde\refto{lnc}
has also used a path integral prescription, which in the simple models that
have been studied so far, gives the same results as the tunneling wave
function.

Of course, these two boundary conditions are not the only possible ones,
but I will concentrate on them because they are relatively well studied.


\beginsection{COSMOLOGICAL PREDICTIONS FROM $\psi$}

To see what kinds of cosmological predictions we can get from different
boundary conditions, I would like to consider a somewhat more realistic
model than the one I discussed at the beginning.  Instead of a constant
vacuum energy $\rho_v$, I introduce a scalar field $\varphi$ with a potential
$V(\varphi)$.  Since
vacuum energy is zero (or very small) in our part of the universe, $V(\varphi)$
should have a minimum with $V=0$.  The WDW equation for this two-dimensional
model can be solved assuming that $V(\varphi)$ is a slowly-varying function,
$|V'/V| << m_{\rm pl}^{-1}$, which is
well below the Planck density, $\rho_{\rm pl} = m^4_{\rm pl}$, for all
values of $\varphi$.  A slowly-varying $V(\varphi)$ helps
to simplify the
equation, but is also necessary for the inflationary scenario.  If the
condition $V(\varphi) << \rho_{\rm pl}$ is violated, then the semiclassical
approximation is not valid
and higher-order corrections to quantum gravity are important.  Potentials used
in particle physics are often unbounded from above; however, one can also
consider sigma-model-type theories in which $\varphi$ is defined on a compact
manifold and $V(\varphi)$ is bounded.  Another possibility would be to allow an
infinite range of $\varphi$ with $V(\varphi)$ unbounded from below, but bounded
from above.

The tunneling boundary condition supplemented by the condition of
regularity, $|\psi| < \infty$, defines a unique wave function $\psi$.  The
corresponding
probability distribution for the initial values of $\varphi$ in nucleating
universes is\refto{avqc}
$$
{\cal P}_T \propto \exp \left( -{3\rho_{\rm pl} \over{8V(\varphi)}}\right) \,.
\eqno (15)
$$
This probability is strongly peaked at the value $\varphi = \varphi_*$ where
$V(\varphi)$ has a maximum.
Thus, the tunneling wave function ``predicts" that the universe nucleates
with the largest possible vacuum energy.  This is just the right initial
condition for inflation.  The high vacuum energy drives the inflationary
expansion, while the field $\varphi$ gradually ``rolls down" the potential
hill,
and ends up at the minimum with $V(\varphi) = 0$, where we are now.  The
predicted
initial size of the universe is $a_{\rm min} = [3/8\pi GV(\varphi_*)]^{1/2}$.

When the same procedure is repeated for the Hartle-Hawking boundary
condition, one finds a probability distribution similar to (15), but with a
crucial difference in sign,
$$
{\cal P}_H \propto \exp \left( +{3\rho_{\rm pl} \over{8V(\varphi)}}\right) \,.
\eqno (16)
$$
This distribution is peaked at $V(\varphi) = 0$, and thus the Hartle-Hawking
wave function
appears to predict an empty universe with $\rho \approx 0$.  Such initial
condition does
not lead to inflation and is therefore inconsistent with observations.

An attempt to rescue the Hartle-Hawking wave function has been made by
Grischuk and Rozhansky\refto{gr}.  They assert that Lorentzian classical
trajectories
for the nucleating universes begin at the caustics formed by a class of
Euclidean trajectories.  Analysis of the caustics shows that they do not
extend into the dangerous region where $V(\varphi)$ is very small, and the
authors
conclude that no trajectories begin in that region of superspace.  I
disagree with this picture.  It is true that the under-barrier evolution can
sometimes be represented as a motion in imaginary time.  When this is possible,
Euclidean and Lorentzian trajectories can be matched only at points where
$\nabla S = 0$, and it is clear from eq.(12) that such points must lie on the
surface $U(a, \varphi) = 0$.  This ``barrier'' does not coincide with the
caustics of Ref. \cite{gr} and extends all the way to $V(\varphi) = 0$.  In a
more general case, the under-barrier action is complex \refto{bg}, and no
conclusion can be reached on the basis of purely Euclidean
trajectories.\footnote{b)}{It should be noted that the assumption of slow
variation of $V(\varphi)$ breaks down near $V(\varphi) = 0$.  Although the
probability ${\cal P}(\varphi)$ is expected to be large in this region, the
analytic approximation (16) cannot be used.}

It appears that  the only escape route for the Hartle-Hawking wave function is
to relax the  condition on $V(\varphi)$ and allow it to reach over-Plankian
values\refto{hp}. It is quite  possible that Planckian densities were reached
or exceeded at the origin of  the universe, but we have no idea how to handle
this case, and cannot make  any predictions.


\beginsection{ETERNAL INFLATION}

The problem of ``eternal inflation'' is not strictly a quantum cosmological
problem, but it is very important for quantum cosmology, and I think it is
appropriate to discuss it here.  Once started, inflation never ends
completely\refto{vet, set, let, avet, lbk}.  Thermalization of the false vacuum
energy is a stochastic
process and does not occur simultaneously in all space.  Regions of false
vacuum constantly undergo thermalization, and these thermalized regions
expand at the speed approaching the speed of light.  However, the inflating
regions between them expand even faster and do not allow them to merge and
fill the entire universe.  Hence, there are large parts of the universe in
which inflation continues even at this time.

This picture suggests the possibility that inflation can also be extended
to the infinite past, avoiding in this way the problem of initial
singularity.  The universe would then be in a steady state of eternal
inflation which does not have a beginning.  It appears, however, that such an
extension is impossible and that eternal inflation must have a beginning in
time.

The impossibility of inflation without a beginning can be easily understood in
the simple de Sitter model (1), (3) with a constant vacuum
energy.  The metric (3) describes contracting and re-expanding universe.
Thermalized regions are driven apart during the expanding phase, but they
would easily merge and fill the entire universe during the contraction.
The universe would then collapse to a singularity without ever getting to
the expanding phase.  Hence, deSitter spacetime cannot be used to describe
eternal inflation.

In more realistic models, the spacetime is locally approximately de Sitter, but
can be globally quite different.  To study the problem in the general case, I
formulated conditions that a spacetime should satisfy in order to describe an
eternally inflating universe. \refto{vil} I was then able to prove that these
conditions cannot be satisfied in a two-dimensional spacetime and gave a
plausibility argument for the more difficult four-dimensional case.

We are thus led to an unexpected and somewhat bizarre conclusion that the
universe had a beginning, but will have no end.  As inflation continues, new
regions of thermalization are formed, and we live in one of these regions.  Our
region is likely to be at a very large (but finite) separation from the
``moment of creation''.


\beginsection{ACKNOWLEGEMENT}

\noindent
{I am grateful to Leonid Grishchuk for discussions.
This work was supported in part by the National Science Foundation.}

\references

\refis{dw} De Witt, B. S. 1967. Phys. Rev. {\bf 160}: 1113

\refis{cm} Misner, C. W. 1972. {\it In} ``Magic Without Magic'', Freeman, San
Francisco

\refis{et} Tryon, E. P. 1973. Nature {\bf 246}: 396

\refis{pif} Fomin, P. I. 1975. Doklady Akad. Nauk Ukr. SSR. {\bf 9A}: 831

\refis{avpl} Vilenkin, A. 1982. Phys. Lett. {\bf 117B}: 25

\refis{gz} Grishchuk, L. P. \& Y. B. Zel'dovich. 1982. {\it In} ``Quantum
Structure of Space and Time''. M. Duff \& C. Isham, Eds. Cambridge University
Press,
      Cambridge.

\refis{hh} Hartle, J. B. \& S. W. Hawking. 1983. Phys. Rev. {\bf D28}: 2960

\refis{lnc} Linde, A. D. 1984. Lett. Nuovo Cimento {\bf 39}: 401

\refis{avpr} Vilenkin, A. 1984. Phys. Rev. {\bf D30}: 509

\refis{avint} Vilenkin, A. 1989. Phys. Rev. {\bf D39}: 1116

\refis{lr} Lapshinsky, V. \& V. A. Rubakov. 1979. Acta Phys. Pol. {\bf B10}:
1041

\refis{halh} Halliwell, J. J. \& S. W. Hawking. 1985. Phys. Rev. {\bf D31}:
1777

\refis{ban} Banks, T. 1985. Nucl. Phys. {\bf B249}: 332

\refis{avbc} Vilenkin, A. 1986. Phys. Rev. {\bf D33}: 3560

\refis{kk} Kuchar, K. 1992. {\it In} Proceedings of the 4th Canadian Conference
on General Relativity and Relativistic Astrophysics. World Scientific.
Singapore.

\refis{ci} Isham, C. 1992. Imperial College Preprint TP/91-92/25

\refis{haw} Hawking, S. W. 1984. Nucl. Phys. {\bf B239}: 257

\refis{harhal} Halliwell \& Hartle

\refis{avqc} Vilenkin, A. 1988. Phys. Rev. {\bf D37}: 888

\refis{gr} Grishchuk, L. P. \& L. Rozhansky. 1988. Phys. Lett. {\bf 208B}: 369

\refis{hp} Hawking, S. W. \& D. N. Page. 1986. Nucl. Phys. {\bf B264}: 185

\refis{vet} Vilenkin, A. 1983. Phys. Rev. {\bf D27}: 2848

\refis{set} Starobinsky, A.A. 1986. {\it In} Field Theory, Quantum Gravity and
Strings, H. J. de Vega and N. Sanchez, Eds. Springer. Berlin.

\refis{let} Linde, A. D. 1986. Mod. Phys. Lett. {\bf A1}: 81

\refis{lbk} Linde, A. D. 1990. Particle Physics and Inflationary Cosmology.
Harwood Academic. Chur.

\refis{vil} Vilenkin, A. 1992. Phys. Rev. {\bf D46}: 2355

\refis{bg} Bowcock, P. \& R. Gregory. 1991. Phys. Rev. {\bf D44}: 1774

\refis{avet} Aryal, M. \& A. Vilenkin. 1987. Phys. Lett. {\bf B199}: 351

\endreferences
\medskip

\vfill
\eject

\endjnl
\end